\documentclass[preprint, aps, preprintnumbers,groupedaddress,nofootinbib]{revtex4-1}

\pdfoutput=1
\usepackage{graphicx}
\usepackage{latexsym}
\usepackage{amsfonts}
\usepackage{amssymb}
\usepackage{amsmath}
\usepackage{slashed}
\usepackage{feynmp}
\usepackage{hyperref}
\usepackage{url}
\usepackage{color}
\usepackage{cancel}
\usepackage{multirow,array}

\usepackage{dcolumn}
\usepackage{bm}

\usepackage[utf8]{inputenc}

\begin{document}

\title{Direct  measurement  of the Higgs self-coupling in $e^+_{}e^-_{} \to ZH$}

\author{Junya Nakamura}
\email{junya.nakamura@itp.uni-tuebingen.de}
\affiliation{Institut f\"ur Theoretische Physik, Universit\"at T\"ubingen, 72076 T\"ubingen, Germany}

\author{Ambresh Shivaji}
\email{ambresh.shivaji@uclouvain.be}
\affiliation{Centre for Cosmology, Particle Physics and Phenomenology (CP3), Universit\'{e} Catholique de Louvain, B-1348 Louvain-la-Neuve, Belgium}

\preprint{CP3-18-71}

\begin{abstract}

A new method to measure  the trilinear Higgs self-coupling $\lambda_3^{}$ in a single Higgs production process is proposed. 
Time-reversal-odd (T-odd) asymmetries in the process $e^+_{}e^-_{} \to ZH$, $Z \to f\bar{f}$ are computed  from the absorptive part of the electroweak one-loop amplitude. 
Since the T-odd asymmetries measure the tree-level $t$-channel $ZH \to ZH$ scattering, they can be direct probes of $\lambda_3^{}$.  
The proposed method  is quite challenging; a relatively large statistics and polarized  $e^+_{}e^-_{}$ beams are demanded. However, this is probably the only approach to directly measure $\lambda_3^{}$ in $e^+_{}e^-_{}$ collisions, when  a beam energy above the $ZHH$ production threshold is not available.

\end{abstract}

\pacs{}

\maketitle


The capabilities of the LHC and future $e^+_{}e^-_{}$ colliders to measure the trilinear Higgs self-coupling $\lambda_3^{}$ have been extensively studied in recent years~\cite{Baur:2002rb, Baur:2003gp, Baur:2009uw, Dolan:2012rv,Baglio:2012np, Baer:2013cma, Asner:2013psa, McCullough:2013rea, Frederix:2014hta,Azatov:2015oxa,Shen:2015pha,Gorbahn:2016uoy, Degrassi:2016wml, Bizon:2016wgr, Bishara:2016kjn, DiVita:2017eyz, DiVita:2017vrr,Maltoni:2017ims, ATL-PHYS-PUB-2017-001,CMS-PAS-HIG-17-030,Goncalves:2018yva,Maltoni:2018ttu, Rindani:2018ubx,ATLAS-CONF-2018-043,Bizon:2018syu,Borowka:2018pxx}. 
Unlike the couplings of the Higgs bosons with heavy fermions and gauge bosons, we do not have any meaningful information on $\lambda_3^{}$ and its 
value can be very different from the one predicted in the standard model (SM). The measurements from the di-Higgs production processes, which are commonly referred as \emph{direct} measurements, are challenging, because of their very small cross sections  both at the LHC~\cite{Baur:2002rb,Baur:2003gp,Dolan:2012rv,Baglio:2012np,Frederix:2014hta,Azatov:2015oxa,Bishara:2016kjn,Goncalves:2018yva} and at $e^+_{}e^-_{}$ colliders~\cite{Baur:2009uw,Baer:2013cma, Asner:2013psa,DiVita:2017vrr,Maltoni:2018ttu},
at which a relatively high beam energy is required. 
Current LHC runs are only able to provide exclusion limits for di-Higgs production~\cite{CMS-PAS-HIG-17-030,ATLAS-CONF-2018-043} and, even the projected sensitivity at 
the HL-LHC is very weak ($-0.8 < \lambda_3/\lambda_3^{\rm SM} < 7.7$)~\cite{ATL-PHYS-PUB-2017-001,Borowka:2018pxx}. 
The information on $\lambda_3^{}$ may be also obtained from  measurement of the (differential) cross sections of  the single Higgs production processes~\cite{McCullough:2013rea, Shen:2015pha, Gorbahn:2016uoy, Degrassi:2016wml, Bizon:2016wgr, DiVita:2017vrr, DiVita:2017eyz, Maltoni:2017ims, Maltoni:2018ttu, Rindani:2018ubx, CMS-PAS-FTR-18-020}. The coupling $\lambda_3^{}$ contributes to the electroweak one-loop correction to single Higgs processes. 
This approach is commonly called an \emph{indirect} one  due to the fact that the trilinear coupling enters into the loop. 
Since   very precise determination of the cross sections is demanded, the indirect method in single Higgs processes is as challenging as the direct measurement from the di-Higgs productions~\cite{McCullough:2013rea, DiVita:2017vrr, DiVita:2017eyz, Maltoni:2018ttu}. 

In $e^+_{}e^-_{}$ collisions, the aforementioned indirect method in the process $e^+_{}e^-_{} \to ZH$ has an obvious advantage over the direct approach with the di-Higgs production processes such as $e^+_{}e^-_{} \to ZHH$; only a smaller beam energy is needed. 
However, the indirect measurements highly depend on assumptions about unknown new physics (NP) at high scale that does not  modify the coupling $\lambda_3^{}$ itself~\cite{McCullough:2013rea}~\footnote{It is known that a virtual heavy fermion in the one-loop radiative correction does not decouple from the cross section measured at low energy~\cite{Fleischer:1982af, Fleischer:1982us, Dawson:1990xt, Kniehl:1991hk}.}. 
Direct methods, on the other hand, are less model dependent and, therefore, can provide more reliable bounds on a possible modification of $\lambda_3^{}$.

In this work, a method of measuring  \emph{directly} the coupling  $\lambda_3^{}$  in the single Higgs production process $e^+_{}e^-_{} \to ZH$ is proposed. The method deals with time-reversal-odd (T-odd) asymmetries in the production process with a subsequent $Z$ boson decay into a massless fermion pair,
\begin{align}
e^+_{}e^-_{} \to Z (\to f \bar{f}) + H. \label{eq:full-process}
\end{align}
When CP (or equally T) is conserved, T-odd quantities are generally identical to zero in the tree-level approximation and receive finite contributions only from an absorptive part of loop diagrams~\cite{DeRujula:1972te}. 
The T-odd asymmetries are computed at the lowest order from the absorptive part of the electroweak one-loop amplitude. 
The absorptive part always includes the tree level $t$-channel $ZH \to ZH$ re-scattering effect, a part of which is proportional to  the coupling  $\lambda_3^{}$. 
As a result, the T-odd asymmetries are direct probes of $\lambda_3^{}$.
Unknown heavy NP particles, which may affect the indirect $\lambda_3^{}$ measurement via one-loop radiative corrections, do not contribute to the T-odd asymmetries unless the beam energy is large enough to directly produce these NP particles, because the asymmetries arise only from  the absorptive part~\footnote{It should be noted that the attempt to directly measure a coupling which enters a process at the next-to-leading order  by using T-odd observables is not new in particle physics; see e.g.~\cite{DeRujula:1978bz, Fabricius:1980wg, Hagiwara:1981qn, Hagiwara:1982cq, Hikasa:1982un}.}.

\begin{figure}[th!]
\includegraphics[scale=0.8]{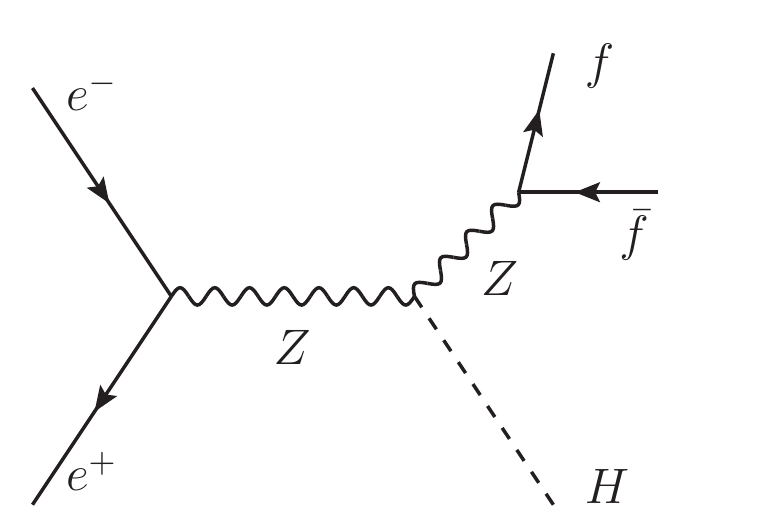}
\caption{Tree-level Feynman diagram for the process $e^+_{}e^-_{} \to Z (\to f \bar{f}) + H$.\label{figure:diagram-tree}}
\end{figure}

The kinematics of the process in Eq.~(\ref{eq:full-process}) can be specified by four independent variables after integration over the azimuthal angle of the $Z$: the center-of-mass (c.m.) energy squared $s$, the polar angle $\Theta$ ($0 \le \Theta \le \pi$) of the $Z$ from the direction of the $e^-_{}$ momentum in the $e^+_{}e^-_{}$ c.m. frame, the polar $\theta$ and azimuthal $\phi$ angles of the final fermion $f$ in the $Z$ rest frame. 
We neglect the initial electron mass and the final fermion mass. 
For given $s$ and the electron helicity ($\tau$), after the summation over the final fermion helicity, the differential cross section  using the narrow width approximation for the $Z$ boson can be expressed as~\footnote{$s$ dependence is always implicit throughout the paper.}
\begin{widetext}
\begin{align}
  \frac{d^3\sigma(\tau)}{d\cos{\Theta} d\cos{\theta}d\phi} & = 
   F_{1}^{} (1+\cos^2_{}{\theta} )
+ F_{2}^{} (1-3\cos^2_{}{\theta} )
+ F_{3}^{} \sin{2\theta} \cos{\phi}    
 + F_{4}^{} \sin^2_{}{\theta} \cos{2\phi} \nonumber  \\
 & + F_{5}^{} \cos{\theta}
+  F_{6}^{}  \sin{\theta} \cos{\phi} 
 +  F_{7}^{} \sin{\theta} \sin{\phi} 
+ F_{8}^{} \sin{2\theta} \sin{\phi} 
+ F_{9}^{} \sin^2_{}{\theta} \sin{2\phi}, 
\label{differential-2}
\end{align}
\end{widetext}
where 
 the nine coefficients $F_{i}^{}$ ($i=1$ to $9$) are functions of $\tau$,  $s$ and $\cos\Theta$. After integrations over $\theta$ and $\phi$, only $F_1^{}$ remains in Eq.~(\ref{differential-2}),
which corresponds to the differential cross section for the $ZH$ production process.

\begin{figure}[th!]
\includegraphics[scale=0.47]{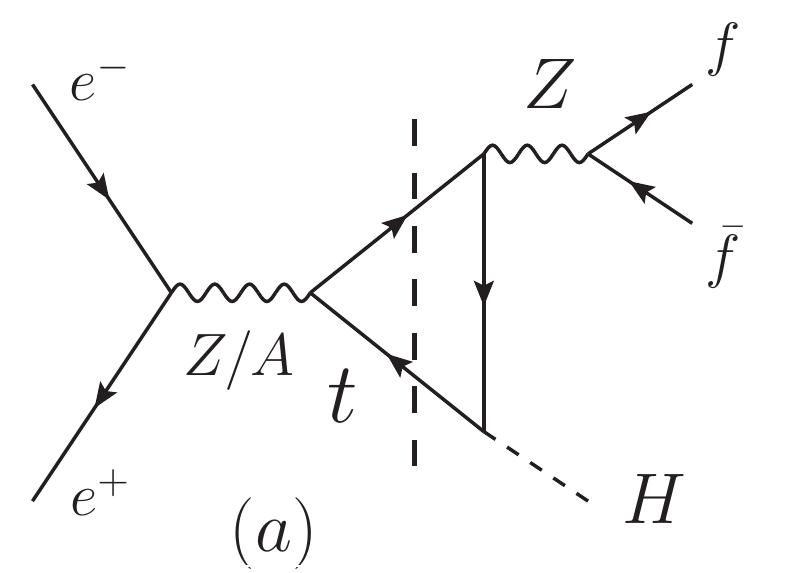}
\includegraphics[scale=0.47]{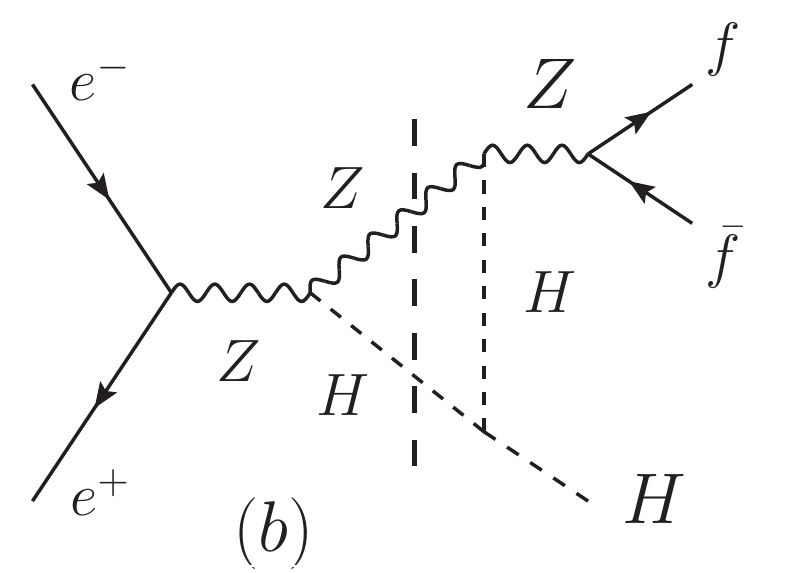}
\includegraphics[scale=0.47]{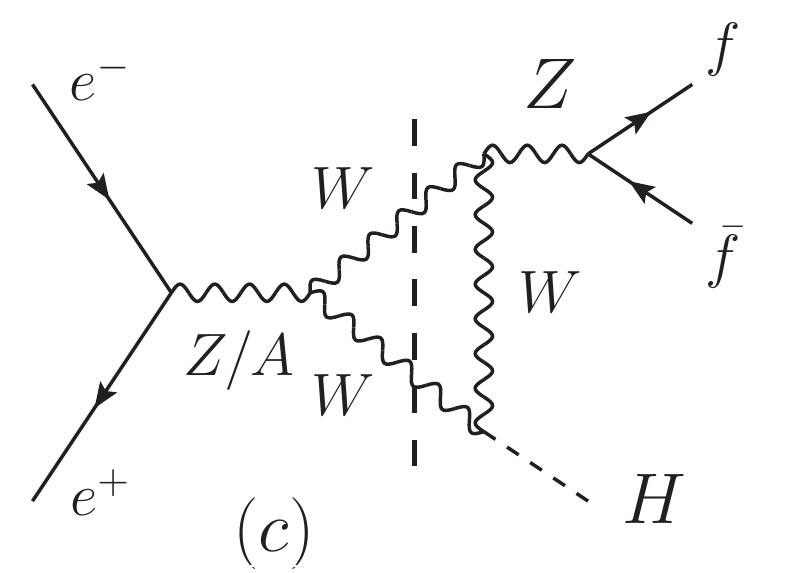}
\includegraphics[scale=0.47]{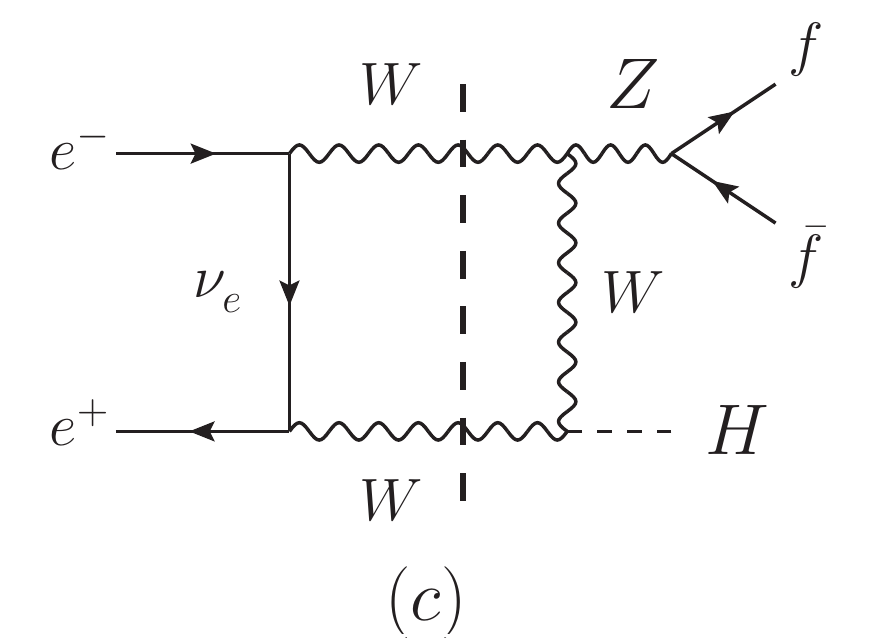}
\caption{Representative one-loop Feynman diagrams that contribute to the T-odd distribution, (a) the top loop diagrams, (b) the Higgs loop diagram that depends on the coupling $\lambda_3^{}$, and (c) a part of the gauge boson loop diagrams. \label{figure:loop-diagram}}
\end{figure}

The first six coefficients $(F_{1}^{}, F_{2}^{}, F_{3}^{}, F_{4}^{}, F_{5}^{}, F_{6}^{})$ are T-even and the last three coefficients $(F_{7}^{}, F_{8}^{}, F_{9}^{})$ are T-odd.  The leading contribution to the six T-even coefficients is calculated from the tree diagram shown in Fig.~\ref{figure:diagram-tree}. 
The amplitude for the production process  can be in general written as
\begin{align}
{\cal M}_{\tau}^{\lambda} = \bar{v}(\bar{p},-\tau) \Gamma^{\mu}_{} u( p,\tau) \epsilon^{*}_{\mu}(k,\lambda), \label{eq:amplitude-1}
\end{align}
where $u$ and $v$ are the spinors for the electron and the positron, respectively,  $\epsilon^{}_{\mu}$ the $Z$ polarization vectors, $\lambda$ the  $Z$ boson helicity, and $p$,  $\bar{p}$ and $k$ are the four-momenta of the electron, positron and $Z$ boson, respectively. 
The four-vector $\Gamma_{\mu}^{}$ in the one-loop calculation can be expanded as
\begin{align}
\Gamma_{\mu}^{} = \sum_{\rho=\mp}^{} \bigl( a_{(1) \rho}^{} \gamma_{\mu}^{}  + a_{(2) \rho}^{} p_{\mu}^{} \slashed{k} + a_{(3) \rho}^{} \bar{p}_{\mu}^{} \slashed{k}  \bigr) \frac{1 + \rho \gamma_5^{}}{2}. \label{eq:expand}
\end{align}
The six coefficients $a_{(i)\mp}^{}$ ($i=1$ to $3$)  are complex numbers, independent of $\tau$ and $\lambda$. 
 In the tree-level calculation, only $a_{(1)\mp}^{}$ are non-zero:
\begin{align}
& A_{-}^{}  \equiv a_{(1)-}^{(\mathrm{tree})}  = (s - m_Z^{2})^{-1}_{} \frac{4\pi \alpha}{s_W^2c_W^2} \bigl(-1/2 + s^2_{W} \bigr) m_Z^{}, \nonumber \\
& A_{+}^{}  \equiv a_{(1)+}^{(\mathrm{tree})}   = (s - m_Z^{2})^{-1}_{} \frac{4\pi \alpha}{c_W^2}  m_Z^{}, \nonumber \\
& a_{(2)-}^{(\mathrm{tree})}   = a_{(2)+}^{(\mathrm{tree})}  = a_{(3)-}^{(\mathrm{tree})}  = a_{(3)+}^{(\mathrm{tree})}  = 0,\label{eq:coefficients-tree}
\end{align} 
where $\alpha = e^2_{}/(4\pi)$ with $e$ being the magnitude of the electron charge,
$c_W^{} = \cos\theta_W^{}$ and $s_W^{} = \sin\theta_W^{}$ are the weak mixing factors. 
We define the coordinate system of the $Z$ rest frame as follows: the $z$ axis is  along the original $Z$ momentum direction and the $y$ axis is along the direction of $\vec{p} \times (-\vec{\bar{p}})$.
The tree-level prediction for the angular coefficients in our coordinate system is
\begin{align}
F_i^{}(\tau,  \cos\Theta) & = 
\frac{3|\vec{k}|}{256\pi^2_{} s^{3/2}_{}}
\sum_{f,\tau^{\prime}_{}}
\mathrm{B}_{f}^{} \frac{(v_f^{}+ \tau^{\prime}_{} a_f^{})^2_{}}{2(v_f^2+a_f^{2})} 
f_i^{}(\tau, \tau^{\prime}_{},  \cos\Theta)\label{eq:TevenF}
\end{align}
with
\begin{align}
f_1^{} & =  A_{\tau}^2 s \Bigl( 1+\cos^2_{}\Theta  +\frac{w^2_{}}{m_Z^2} \sin^2_{}\Theta \Bigr), \nonumber \\
f_2^{} & =  A_{\tau}^2 s \frac{w^2_{}}{m_Z^2} \sin^2_{}\Theta , \nonumber \\
f_3^{} & = - A_{\tau}^2 s \frac{w}{m_Z^{}} \sin2\Theta , \nonumber \\
f_4^{} & = A_{\tau}^2 s \sin^2_{}\Theta, \nonumber \\ 
f_5^{} & = 4 \tau \tau^{\prime}_{} A_{\tau}^2 s \cos\Theta , \nonumber \\
f_6^{} & = - 4 \tau \tau^{\prime}_{} A_{\tau}^2 s \frac{w}{m_Z^{}} \sin\Theta, \nonumber \\
f_7^{} & = f_8^{}  = f_9^{}  = 0, \label{eq:Tevencoefficients}
\end{align}
where $w$ is the $Z$ boson energy $w = (s + m_Z^2 - m_H^2)/(2\sqrt{s})$, $\vec{k}$ is the $Z$ boson three momentum $|\vec{k}| = \sqrt{w^2_{}-m_Z^2}$, $\mathrm{B}_f^{}$ is the branching fraction $\mathrm{B}_f^{} = \Gamma(Z \to f\bar{f})/\Gamma(Z \to \mathrm{all})$, $v_f^{}$ and $a_f^{}$ are the vector and axial-vector couplings of the $Z$ to the final fermion $f$, and $\tau^{\prime}_{}$ is the final fermion helicity. 
At the tree-level, the T-odd coefficients are vanishing as expected. 
The first non-vanishing contribution comes from the interference between the tree diagram and the absorptive part of the one-loop diagrams~\footnote{Here the $Z \to f\bar{f}$ decay is always the tree level calculation. This is sufficient for our purpose, since the absorptive part of the one-loop diagrams in $Z \to f\bar{f}$ does not produce the T-odd distribution in the massless fermion limit.}:
\begin{align}
 f_7^{}  & =  2 \tau^{\prime}_{} A_{\tau}^{} \frac{s^{3/2}_{}}{m_Z^{}} |\vec{k}| \sin\Theta  \Bigl\{ w \mathrm{Im}\Bigl( a_{(2)\tau}^{(\mathrm{loop})} - a_{(3)\tau}^{(\mathrm{loop})}  \Bigr) - |\vec{k}| \cos\Theta \mathrm{Im}\Bigl( a_{(2)\tau}^{(\mathrm{loop})} + a_{(3)\tau}^{(\mathrm{loop})} \Bigr) \Bigr\}, \nonumber \\
 f_8^{}  & =  \tau A_{\tau}^{} \frac{s^{3/2}_{}}{m_Z^{}} |\vec{k}| \sin\Theta \Bigl\{ w \cos\Theta \mathrm{Im}\Bigl( a_{(2)\tau}^{(\mathrm{loop})} - a_{(3)\tau}^{(\mathrm{loop})}  \Bigr) - |\vec{k}| \mathrm{Im}\Bigl( a_{(2)\tau}^{(\mathrm{loop})} + a_{(3)\tau}^{(\mathrm{loop})}  \Bigr)  \Bigr\}, \nonumber \\
 f_9^{}  & =  - \tau A_{\tau}^{} s^{3/2}_{} |\vec{k}| \sin^2\Theta \mathrm{Im}\Bigl( a_{(2)\tau}^{(\mathrm{loop})} - a_{(3)\tau}^{(\mathrm{loop})}  \Bigr) .\label{eq:Toddcoefficients}
\end{align}
Note that the absorptive part in this order is both ultraviolet and infrared finite.
We notice that $\mathrm{Im}( a_{(1)\mp}^{(\mathrm{loop})})$ do not contribute to the T-odd coefficients. 
As a result, we need to calculate the absorptive part of only a limited one-loop diagrams, representatives of which are shown in Fig.~\ref{figure:loop-diagram}. 
We divide the relevant one-loop diagrams   
into three categories, namely, top loop diagrams, a Higgs loop diagram that depends on the coupling $\lambda_3^{}$, and gauge boson loop diagrams. These are labeled as (a), (b) and (c), respectively, in Fig.~\ref{figure:loop-diagram}.  
They are separately gauge-independent. 
The electroweak one-loop diagrams and amplitude are generated with help of {\tt FeynArts}~\cite{Hahn:2000kx} and {\tt FormCalc}~\cite{Hahn:1998yk}.
The analytic formulas for $\mathrm{Im}( a_{(2)\mp}^{(\mathrm{loop})} )$ and $\mathrm{Im}( a_{(3)\mp}^{(\mathrm{loop})} )$ have been obtained but they are very long expressions and will be provided elsewhere~\cite{nakamura:2018}. The numerical values for the one-loop scalar functions are calculated with the {\tt LoopTools}~\cite{vanOldenborgh:1989wn, Hahn:1998yk}. Phase space integration is performed with {\tt BASES}~\cite{Kawabata:1995th}. 
Our calculation has been numerically checked in the following two ways.
First, CP invariance of the differential cross section~\cite{Hagiwara:1993sw, Hagiwara:2000tk} has been tested. 
Second, the T-odd coefficients  have been also calculated from the electroweak full one-loop helicity amplitudes using {\tt MadGraph5\verb|_|aMC@NLO}~\cite{Alwall:2014hca, Frederix:2018nkq}. We have found perfect agreement  for several phase space points.

\begin{figure}[th!]
\includegraphics[scale=0.5]{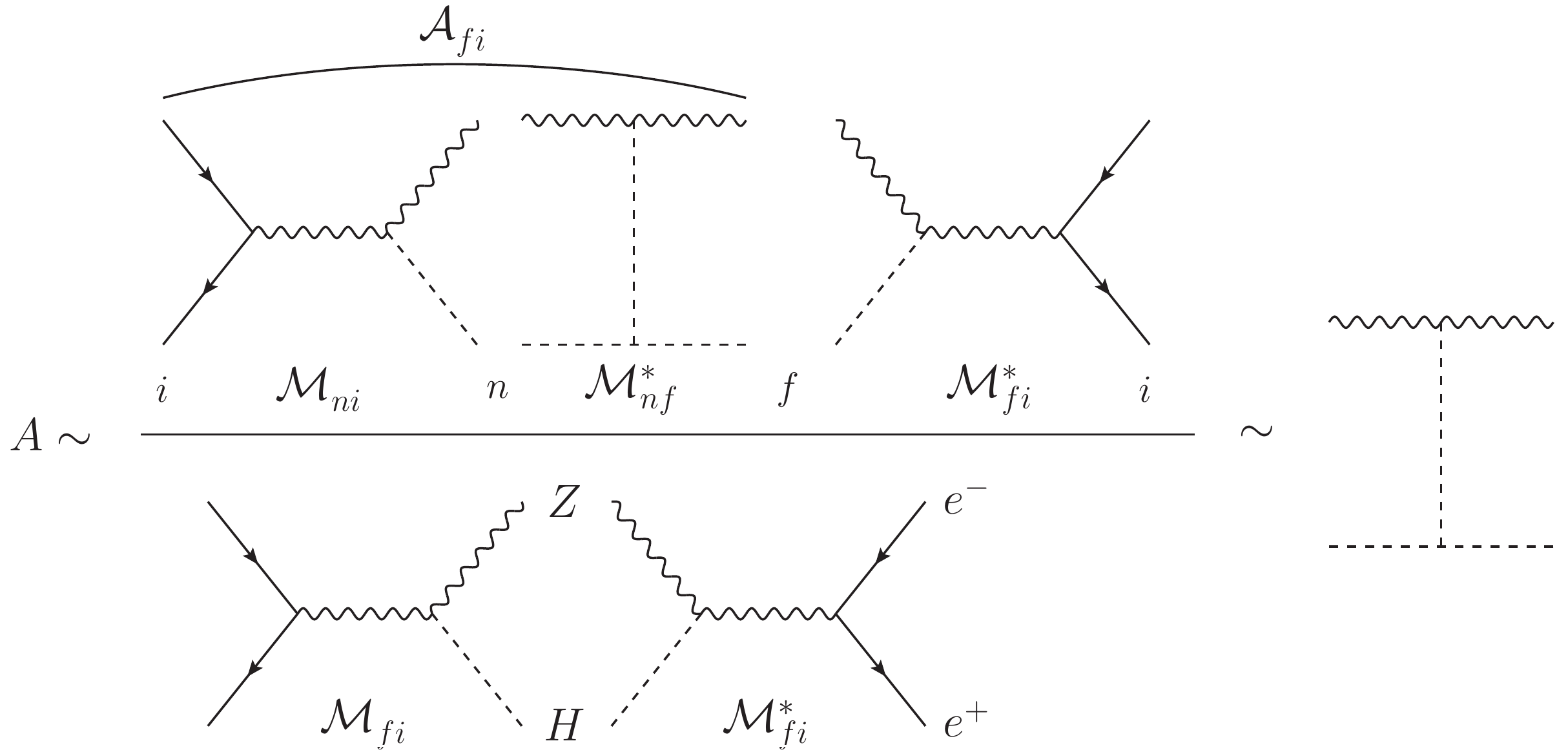}
\caption{Diagrams contributing to the numerator and denominator in the T-odd asymmetries in Eq.~(\ref{eq:t-odd-asymmetries-3}). ${\cal A}_{fi}^{}$ represents the absorptive part of the Higgs one-loop diagram. The asymmetries measure the tree-level $t$-channel $ZH \to ZH$ scattering.     \label{figure:asym-diagrams}}
\end{figure}

The leading order prediction for T-odd asymmetries is of order $g^2_{}/(4\pi)$ and can be obtained by dividing the T-odd coefficients $(F_{7}^{}, F_{8}^{}, F_{9}^{})$  in Eq.~(\ref{eq:Toddcoefficients}) by the T-even coefficient $F_{1}^{}$ in Eq.~(\ref{eq:Tevencoefficients}) . 
We define the integrated T-odd asymmetries by 
\begin{align}
A_7^{} & \equiv \frac{\sum_{\tau} \xi(\tau) \Bigl(\int^1_{0} - \int^0_{-1}  \Bigr) d\cos\Theta F_7^{}(\tau, \cos\Theta)}{\sum_{\tau} \xi(\tau) \int^1_{-1} d\cos\Theta F_1^{}(\tau, \cos\Theta)}, \nonumber \\
A_8^{} & \equiv \frac{\sum_{\tau} \xi(\tau) \int^1_{-1}  d\cos\Theta F_8^{}(\tau, \cos\Theta)}{\sum_{\tau} \xi(\tau) \int^1_{-1} d\cos\Theta F_1^{}(\tau, \cos\Theta)},
\label{eq:t-odd-asymmetries-3}
\end{align}
where $\xi(\tau)=( 1 + \tau P_{e^-_{}}^{} ) ( 1 - \tau P_{e^+_{}}^{} )$ and, $1 \leq P_{e^-_{}}^{}(P_{e^+}) \leq 1$  denotes the degree of longitudinal polarization of the electron (positron) beam. The integration over $\cos\Theta$ takes into account the CP invariance of the differential cross section. 
We have found that the coefficient $F_9^{}$ receives contribution only from the gauge boson loop diagrams (c) in Fig.~\ref{figure:loop-diagram}~\cite{nakamura:2018}, therefore an asymmetry based on it is not useful for our purpose. 
In Fig.~(\ref{figure:asym-diagrams}), diagrams contributing to the numerator and denominator in the T-odd asymmetries are described. Here ${\cal A}_{fi}^{}$ represents the absorptive part of only the Higgs one-loop diagram~\footnote{The explicit formula of T-odd quantities in terms of an absorptive part can be found in Refs.~\cite{DeRujula:1972te, DeRujula:1978bz, Hagiwara:1982cq}.}.   
It is shown that, because an absorptive part of one-loop amplitude is simply a tree amplitude times a tree amplitude, the tree diagram for $e^+_{}e^-_{} \to ZH$ drops from the ratio and only  the tree diagram for the $ZH \to ZH$ scattering ($t$-channel) is left.  
This explains that the T-odd asymmetries measure the tree-level $ZH \to ZH$ scattering, and because the coupling $\lambda_3^{}$ is no longer a  part of the loop, the T-odd asymmetries are direct probes of $\lambda_3^{}$.
The asymmetries depend also on the $ZZH$ coupling. However, the 
$ZZH$ coupling can be constrained separately via a precise measurement 
of the cross section. Therefore, we can use it as 
input to predict the asymmetries, focusing only on constraining deviation in $\lambda_3^{}$.

We use the following set of input parameters for the numerical results:
\begin{align}
m_H^{} = 125,\  m_W^{} = 80.419,\ m_Z^{} = 91.188,\ m_t^{} = 174.3,\nonumber \\
\ 
v = 246.218,\ 
s_W^2=0.222,\ \alpha = 1/132.507  \nonumber 
\end{align}
in units of GeV for the mass parameters.  In the following numerical studies, we restrict ourselves to the case $m_Z^{}+m_H^{} < \sqrt{s} < m_Z^{}+2m_H^{}$, in which case the direct $\lambda_3^{}$ measurement in $e^+_{}e^-_{} \to ZHH$ is not possible.  
Since the c.m. energy chosen is also below the $t\bar{t}$ threshold, i.e. $\sqrt{s} < 2m_t^{}$, the top loop diagrams do not contribute to the T-odd asymmetries and we can avoid any ambiguity from a modified top Yukawa coupling due to a high scale NP, which is unavoidable in the indirect method~\cite{Shen:2015pha}. 
We separate the contribution from the Higgs loop diagram and that from the gauge boson loop diagrams to the SM asymmetries as,
\begin{align}
A_i^{\mathrm{SM}} = A_i^{\mathrm{Higgs}} + A_i^{\mathrm{Gauge}}\ \ \mathrm{for}\ i=7, 8. \label{eq:separate}
\end{align}
Observation of $A_7^{}$ requires the charge identification of the final fermion $f$. 
This requirement is easily met for the decay modes $Z \to \ell^-_{} \ell^+_{}$. 
The charge of a $B$ meson containing one $b$ or $\bar{b}$ quark can be
identified via the decay mode $B \to \ell \nu + X$. 
We assume an efficiency of $20\%$ for identifying the charges of the decaying $b$ or $\bar{b}$ hadrons~\cite{Hagiwara:2000tk}. 
The asymmetries receive unpleasant  suppressions due to the fact that the T-odd coefficients  are also parity-odd 
 in case we do not measure the spin of the initial and final states~\cite{Hagiwara:1982cq}.  
Techniques to reduce the suppressions are as follows. 
The asymmetry $A_7^{}$ vanishes if the $Z \to f\bar{f}$ decay process conserves parity. Since the coupling of the charged lepton to the $Z$ is almost axial-vector, $A_7^{}$ in the $Z \to e^-_{}e^+_{}$ and $Z \to \mu^-_{}\mu^+_{}$ decay modes has the suppression factor of $\sim 1/5$, which is unavoidable. However, for some fraction of  the $Z \to \tau^-_{}\tau^+_{}$ decay,  we can measure the $\tau$ helicity from $\tau$ decay distributions~\cite{Bullock:1992yt, Hagiwara:2012vz}  and reduce the suppression factor. We assume an efficiency of $40\%$ for measuring the $\tau$ helicity~\cite{Hagiwara:2000tk}. 
Similarly, the asymmetry $A_8^{}$ vanishes if parity is conserved in the production process, namely if $a_{(i)-}^{}=a_{(i)+}^{}$ for all $i=1,2,3$. Due to the coupling of the incoming electron to the $Z$ being dominantly axial-vector, the Higgs loop contribution $A_8^{\mathrm{Higgs}}$ is very suppressed without beam polarization. Fortunately, polarized beams can be available in future $e^-_{}e^+{}$ colliders~\cite{Baer:2013cma}. 

\begin{figure}[th!]
\centering
\includegraphics[scale=0.9]{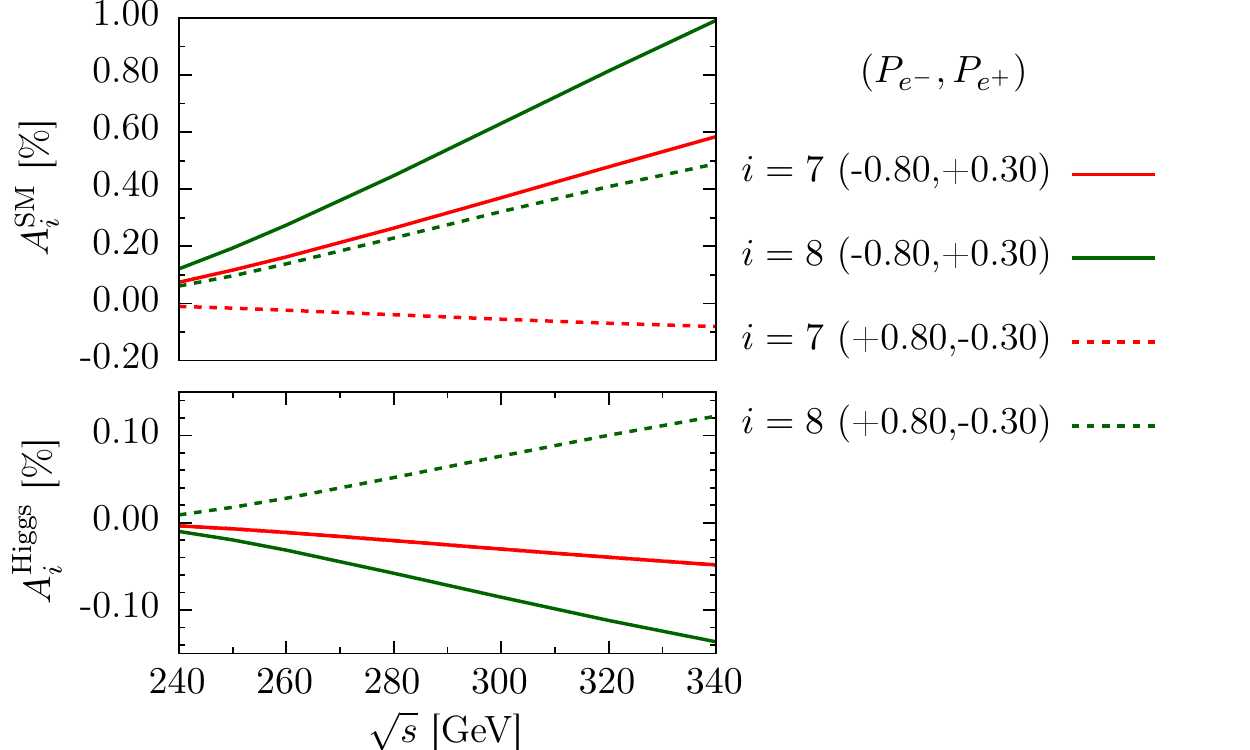}
\caption{$(A_7^{\mathrm{SM}}, A_8^{\mathrm{SM}})$ in the upper plot and$(A_7^{\mathrm{Higgs}}, A_8^{\mathrm{Higgs}})$ in the lower plot as functions of the c.m. energy, for two choices 
of beam polarizations:  $(-0.80,+0.30)$ (solid) and $(+0.80,-0.30)$ (dashed). 
$A_7^{\mathrm{Higgs}}$ for the two choices 
of polarizations are degenerate. 
 \label{figure:asymmetries}}
\end{figure}

In Fig.~\ref{figure:asymmetries}, we display the variation of 
$(A_7^{\mathrm{SM}}, A_8^{\mathrm{SM}})$ and $(A_7^{\mathrm{Higgs}}, A_8^{\mathrm{Higgs}})$  as functions of the c.m. energy for two practical choices 
of beam polarizations:  $(P_{e^-}^{}, P_{e^+}^{})=(-0.80,+0.30)$, and $(+0.80,-0.30)$. We find that the absolute values of the asymmetries 
are small ($\lesssim 1\%$) and they grow as the beam energy is increased. For both the choices 
of beam polarizations, $(A_8^{\rm SM}, A_8^{\rm Higgs})$ are larger than $(A_7^{\rm SM}, A_7^{\rm Higgs})$, respectively. 
This indicates that $A_8^{}$ can play a more important role than $A_7^{}$ in bounding $\lambda_3^{}$.  
The Higgs loop contribution $(A_7^{\mathrm{Higgs}}, A_8^{\mathrm{Higgs}})$ to the asymmetries also become larger at higher c.m. energies, implying a higher sensitivity to $\lambda_3^{}$ at a higher beam energy.

We parametrize the NP effect on the trilinear self-coupling in terms of a 
real parameter $\delta_h^{}$ as
\begin{align}
\lambda_3^{} =  \lambda_3^{\rm SM} ( 1 + \delta_h^{} ),
\end{align}
where, $\delta_h^{}=0$ gives the SM prediction for $\lambda_3^{}$.  Because $A_i^{\mathrm{Higgs}}$ is proportional to $\lambda_3^{}$, the asymmetries with nonzero $\delta_h^{}$ can be described as
\begin{align}
A_i^{\mathrm{BSM}} = \delta_h^{} \times A_i^{\mathrm{Higgs}} + A_i^{\mathrm{SM}}\ \ \mathrm{for}\ i=7, 8. \label{eq:separate}
\end{align}

\begin{figure}[th!]
\centering
\includegraphics[scale=1.0]{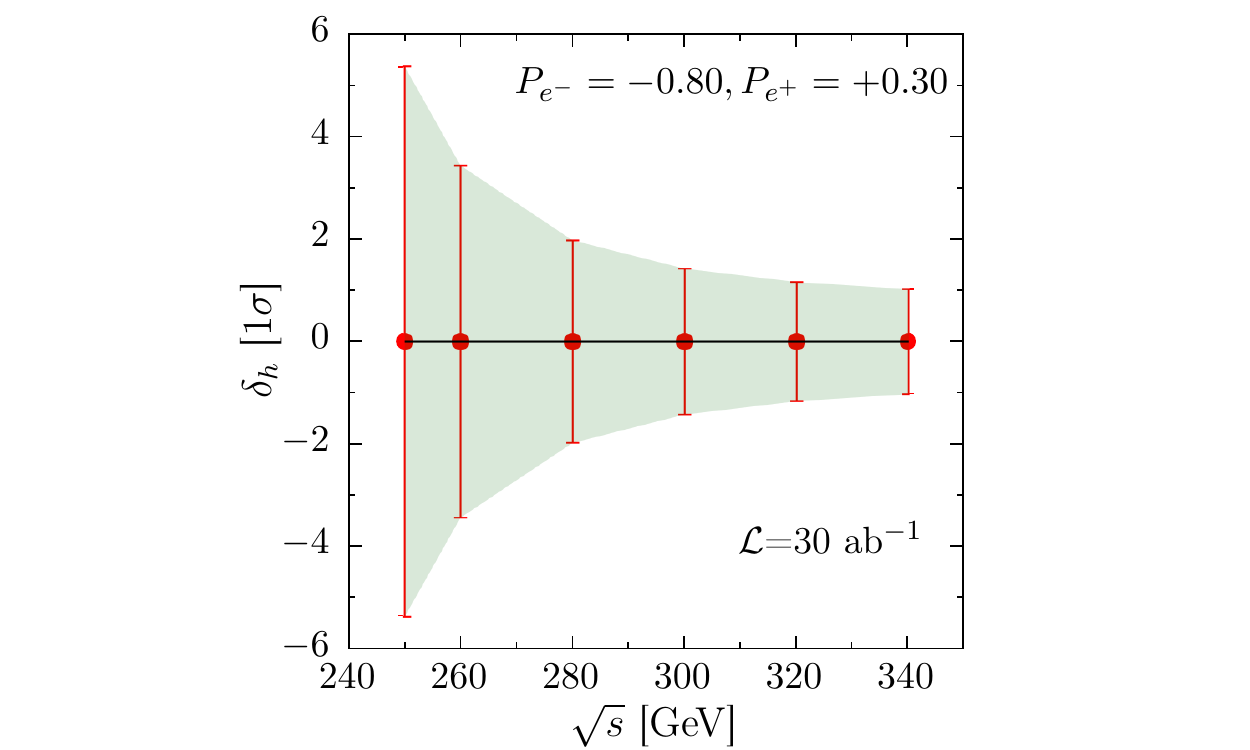}
\caption{Direct $1\sigma$ constraint on $\delta_h^{}$ obtained from the T-odd asymmetries $(A_{7}^{}, A_{8}^{})$ at several c.m. energies, with  beam polarization $(-0.80,+0.30)$ and an integrated luminosity of $30~ \mathrm{ab}^{-1}_{}$.  \label{figure:bound}}
\end{figure}

In Fig.~\ref{figure:bound}, we provide  direct $1\sigma$ bound on 
the SM value of $\delta_h^{}$ that 
can be reached at different c.m. energies by measuring the
asymmetries $(A_{7}^{}, A_{8}^{})$. 
The result is obtained for 
beam polarization of $(-0.80,+0.30)$ which provides better sensitivity due to a larger statistics and the larger asymmetry $A_8^{\mathrm{Higgs}}$; see Fig.~\ref{figure:asymmetries}. 
We have also assumed an integrated luminosity of 30 ab$^{-1}$. By definition, the asymmetries are less sensitive to systematic uncertainties, therefore, in our analysis, we consider only the statistical uncertainty. 
Despite a decrease in the total cross section with the rise of the beam energy, the constraint on $\delta_h^{}$ improves and we can reach an accuracy of about 100\% below the $ZHH$ threshold. 
We have explicitly verified that the accuracy on $\lambda_3$ does not change 
appreciably if uncertainty on $ZZH$ coupling is less than 10\%.
Since the asymmetries depend on $\delta_h^{}$ linearly, the same level of accuracy can be achieved for non-SM values of the parameter $\delta_h^{}$ as well.

To summarize, a direct measurement of the Higgs self-coupling $\lambda_3^{}$ is possible even in the single Higgs production process 
$e^+e^- \to ZH$ by using the T-odd asymmetries.  
Due to the smallness of the asymmetries ($\lesssim 1\%$), the method is very challenging and requires a huge statistics.
Our analysis with a beam polarization $(-0.80,+0.30)$ and an integrated luminosity of 30 ab$^{-1}$ suggests that using this method we can measure $\lambda_3^{}$ with 
an accuracy of $\sim100\%$ at $\sqrt{s}=340$ GeV.
However, the following benefits of the proposed method should be emphasized:

(1) Any ambiguity from a possible modification in the top Yukawa coupling is absent in a measured $\lambda_3^{}$, when the beam energy is below the $t\bar t$ threshold.

(2) Since the T-odd asymmetries are independent information from the $ZH$ production cross section, the $ZZH$ coupling which also contributes to the asymmetries can be very well constrained through the cross section measurement and, therefore, the asymmetries can be utilized to constrain $\lambda_3^{}$ only.

(3) This is so far the only approach to directly measure $\lambda_3^{}$ in $e^+_{}e^-_{}$ collisions, when  a beam energy above the $ZHH$ threshold is not available.\\

\begin{acknowledgments}

The authors wish to thank F. Maltoni and X. Zhao for reading the manuscript. 
The authors are grateful to J. Baglio, D. Gon\c{c}alves, K.~Hagiwara, B.~J\"ager, F. Maltoni, L. D. Ninh and X. Zhao for fruitfull discussions, and V. Hirschi and O. Mattelaer for their help with {\tt MadGraph5\_aMC@NLO} package.
J.N. would also like to thank  K.~Hagiwara for encouragement. 
J.N. would like to acknowledge the warm hospitality of the KEK Theory Center and CP3 Louvain, where he has carried out a part of the work. J.N. is also grateful for the support from MEXT KAKENHI Grant Number JP16K21730.
J.N. very much appreciates the support from the 
Alexander von Humboldt Foundation. The work of A.S. is supported by MOVE-
IN Louvain incoming postdoctoral fellowship co-funded
by the Marie Curie Actions of the European Commission.
\end{acknowledgments}

\bibliography{article}

\end{document}